# The relationship between the radio core-dominance parameter and spectral index in different classes of extragalactic radio sources ($III$)

Zhi-Yuan Pei[1,2,3,4], Jun-Hui Fan[y1,2], Denis Bastieri[1,3,4], Jiang-He Yang[1,5] and Hu-Bing Xiao[1,2,3,4]

[1] Center for Astrophysics, Guangzhou University, Guangzhou 510006, China; *mattpui@e.gzhu.edu.cn*

[2] Astronomy Science and Technology Research Laboratory of Department of Education of Guangdong Province, Guangzhou 510006, China

[3] Dipartimento di Fisica e Astronomia "G. Galilei", Universita` di Padova, I-35131 Padova, Italy

[4] Istituto Nazionale di Fisica Nucleare, Sezione di Padova, I-35131 Padova, Italy

[5] Department of Physics and Electronics Science, Hunan University of Arts and Science,Changde 415000, China

**Abstract** Active galactic nuclei (AGNs) can be divided into two major classes, namely radio-loud and radio-quiet AGNs. A small subset of the radio-loud AGNs is called blazars, which are believed to be unified with Fanaroff-Riley type I and type II (FRI&II) radio galaxies. Following our previous work, we present a latest sample of 966 sources with measured radio flux densities of the core and extended components. The sample includes 83 BL Lacs, 473 FSRQs, 101 Seyferts, 245 galaxies, 52 FRIs&IIs and 12 unidentified sources. We then calculate the radio core-dominance parameters and spectral indices and study their relationship. Our analysis shows that the core-dominance parameters and spectral indices are quite different for different types of sources. We also corroborate that the correlation between core-dominance parameter and radio spectral index extends over all the sources in a large sample presented.

## 1 INTRODUCTION

Active galactic nuclei (AGNs) are interesting and attractive extragalactic sources. Understanding these objects requires extensive knowledge in many different areas: accretion disks, the physics of dust and ionized gas, astronomical spectroscopy, star formation, and the cosmological evolution of galaxies and super massive black holes. The gravitational potential of the supermassive black hole at the center of AGNs is believed to be the ultimate energy source of the AGNs. Roughly, $85\%$ are radio-quiet AGNs, with radio loudness ratio defined as $B = \log \frac{F_{5\,GHz}}{F_{2500\mathring{A}}} < 1.0$ and the remaining ~ 15% are radio-loud AGNs (Fan2005), which have relative strong radio emissions by contradistinguishing with their optical emissions and it is believed to host relativistic jets launched from the central black hole (Urry & Padovani1995).

[5] Corresponding authors: fjh@gzhu.edu.cn; denis.bastieri@unipd.it

Blazars, an extreme subclass of AGNs, are characterized by superluminal motions in their radio components, broadband emissions (radio through $\gamma$-ray), high and variable luminosity, high polarization and core-dominated nonthermal continua etc. (see Abdo et al.2010;Aller et al.2003;Andruchow et al.2005; Cellone et al.2007;Fan et al.1997;Fan2005;Lin & Fan2018;Romero et al.2000,2002;Wills et al.1992; Xie et al.2005;Yang et al.2018a,b;Zhang & Fan2008;Pei et al.2019). These extreme observational properties are believed to be produced by the relativistic Doppler beaming effect on account of a quite small viewing angle between the relativistic jet and the line of sight (Urry & Padovani1995).

Based on the observational properties, there are two subclasses of blazars, namely flat spectrum radio quasar (FSRQ) and BL Lacertae object (BL Lac), with the former displaying strong emission line features and the latter showing very weak or no emission line at all. The two subclasses share many similar observational properties, but their difference in emission lines cannot be ignored.Fan(2003) found that the intrinsic ratio, $f$, which are defined as the comoving luminosity in the jet to the unbeamed luminosity, in BL Lacs is greater than that in FSRQs. The standard model of AGNs foretells that the Fanaroff-Riley type I radio galaxies (hereafter FRIs) are the same parent population of BL Lacs while the parent population ofFSRQs are Fanaroff-Riley type II radio galaxies (hereafter FRIIs) (Fanaroff & Riley1974;Urry & Padovani1995; Fan et al.2011;Pei et al.2019). However, the nature of the central engine of blazars and other classes of AGNs is still an open problem.

In a relativistic beaming model, the emissions are composed of two components (Urry & Shafer1984), namely, the beamed and the unbeamed ones, or core boosted and isotropic extended ones. The observed total emission, $S^{ob}$, is the sum of the beamed, $S^{ob}_{core}$, and unbeamed, $S_{ext.}$ emission, then we have $S^{ob} = S_{ext.} + S^{ob}_{core} = (1 + f\delta^p)S_{ext.}$, where $f = \frac{S^{in}_{core}}{S_{ext.}}$, defined by the intrinsic flux density in the jet to the extend flux density in the co-moving frame, $S^{in}_{core}$is the de-beamed emission in the co-moving frame, $\delta$ is a Doppler factor, and $p$ is depended on the geometrical structure of jet, $p = \alpha + 2$ or $p = \alpha + 3$ refers for a continuous or a moving sphere jet, respectively, and $\alpha$ is the spectral index ($S_\nu \propto \nu^{-\alpha}$). The ratio, $R$, of the two components is the core-dominance parameter (Orr & Browne1982). Some authors adopt the ratio of flux densities while others use the ratio of luminosities to quantify the parameter. Namely, $R = S_{core}/S_{ext.}$ or $R = L_{core}/L_{ext.}$, where $S_{core}$ or $L_{core}$ stands for core emissions while $S_{ext.}$ or $L_{ext.}$ for extended emissions (see Fan & Zhang2003;Fan et al.2011;Pei et al.2016,2019and references therein).

After the launching of *Fermi* Large Area Telescope (LAT), many sources are detected to be the high energy $\gamma$-ray emitters which provides us with a wonderful opportunity to probe the $\gamma$-ray mechanism and extreme properties of AGNs. Based on the first eight years of science data from the *Fermi* Gamma-ray Space Telescope mission, the latest catalog, 4FGL, the fourth *Fermi* Large Area Telescope catalog of high-energy $\gamma$-ray sources, has been released, which includes 5098 sources above in the significance of $4\sigma$, covering from 50 MeV−1 TeV range (The Fermi-LAT collaboration2019a,b). AGNs occupy the vast majority in 4FGL, which 3009 AGNs are included. It comprises 2938 blazars, 38 radio galaxies and 33 other AGNs.

For the blazar sample, it includes 681 flat-spectrum radio quasars (FSRQs), 1102 BL Lac-type objects (BL Lacs) and 1152 blazar candidates of unknown type (BCUs) (The Fermi-LAT collaboration2019a).

In 2011, we compiled a sample of 1223 extragalactic radio sources and calculated their core-dominance parameters ($\log R$) and radio spectral indices ($\alpha_R$). We made the comparison of $\log R$ among BL Lacs, FSRQs, Seyfert galaxies and normal galaxies, FRIs and FRIIs, and particularly, we found this parameter of blazars are on average higher than that of the other subclasses of AGNs. Then we investigated the correlation between $\log R$ and $\alpha_R$ (Fan et al.2011). Previously, we collect a larger catalog contained 2400 radio sources with available core-dominance parameters, and those sources are not listed inFan et al.(2011). We also make the discussion of core-dominated AGNs and related correlation analysis (Pei et al.2019). We obtain the similar conclusions withFan et al.(2011).

In this work, followingFan et al.(2011) andPei et al.(2019), we collect a new sample of radio sources, which are neither included inFan et al.(2011) nor inPei et al.(2019), in total 966 AGNs are shortlisted by cross checked. We then calculate the core-dominance parameters and radio spectral indices, and probe the correlation among them. We enlarge the AGNs sample regarding to the the available core-dominance parameters $\log R$ and make further discussions and re-examine the conclusions drawn in our previous work. Our data are taken from the NASA/IPAC EXTRAGALACTIC DATABASE[1], SIMBAD Astronomical Database[2] and Roma BZCAT[3], from these, we calculate the core-dominance parameters and spectral indices of 966 sources with available radio data. In Section 2, we will present the results; some discussions are given in Section 3. We then conclude and summarize our findings in the final section.

Throughout this paper, without loss of generality, we take $\Lambda CDM$ model, with $\Omega_\Lambda$ c 0.73, $\Omega_M$ c 0.27, and $H_0$ c 73 km s$^{-1}$ Mpc$^{-1}$.

## 2 SAMPLE AND RESULTS

### 2.1 Sample and Calculations

In order to calculate the radio core-dominance parameter and discuss its property, we compiled a list of relevant data from the literature. As a rule, the observations are performed at different frequencies by various authors and studies, nevertheless, most of these data are at 5 GHz, we therefore convert the data given in the literature at other frequencies ($\nu$), to 5 GHz by taking the assumption that (Fan et al.2011;Pei et al.2016, 2019)

$$S^{5\ \mathrm{GHz}}_{\mathrm{core}} = S^{\nu,\mathrm{obs}}_{\mathrm{core}}, \ S^{5\ \mathrm{GHz}}_{\mathrm{ext.}} = S^{\nu,\mathrm{obs}}_{\mathrm{ext.}} (\frac{\nu}{5\ \mathrm{GHz}})^{\alpha_{\mathrm{ext.}}}, \tag{1}$$

then the flux densities are K-corrected, and the core-dominance parameters are finally calculated by using the equation

$$R = (\frac{S_{\mathrm{core}}}{S_{\mathrm{ext.}}})(1+z)^{\alpha_{\mathrm{core}} - \alpha_{\mathrm{ext.}}}, \tag{2}$$

In our calculation, we adopted $\alpha_{\mathrm{ext.}}$ (or $\alpha_{\mathrm{unb}}$) = 0.75 and $\alpha_{\mathrm{core}}$ (or $\alpha_j$) = 0 (Fan et al.2011;Pei et al.2019). For those data given in luminosities, we calculated the core-dominance parameter as $\log R = \log \frac{L_{\mathrm{core}}}{L_{\mathrm{ext.}}}$

---

[1] http://ned.ipac.caltech.edu/forms/byname.html
[2] http://simbad.u-strasbg.fr/simbad/
[3] http://www.asdc.asi.it/bzcat/

and some of them where luminosities that we transform, if necessary, at 5 GHz. For data given in flux densities, we also calculate the luminosity by adopting $L_\nu = 4\pi d_L^2 S_\nu$, where $d_L$ is a luminosity distance, defined as $d_L = (1+z)\dfrac{c}{H_0}\displaystyle\int_1^{1+z}\dfrac{1}{\sqrt{\Omega_M x^3 + 1 - \Omega_M}}\,dx$. In our sample, the sources are mainly obtained at 5 GHz and 1.4 GHz from the literature, we then calculated the spectral indices, $\alpha$ where $S_\nu \propto \nu^{-\alpha}$. If a source has no measured redshift, then the average value of the corresponding group was adopted, only to be used to calculate the core-dominance parameters and luminosities. We evaluate the characteristics of 966 sources and checked their identification according to NED , Roma BZCAT and 4FGL from *Fermi*/LAT. Firstly, for those sources have no identification in NED, they are labeled as "unidentified". In addition, we have 6 sources in our catalog are classified as BCU (blazar candidates of unknown type), which are 0829+275, 1111+4820, 1241+735, 1325−558, 1646−506, 1716−496. We conclude them into sources labeled as "unidentified" as well.

Therefore, our literature survey found 966 extragalactic radio sources, which comprise of 83 BL Lacs, 473 FSRQs, 101 Seyfert galaxies, 52 FRIs & FRIIs, 245 other galaxies, and 12 unidentified sources. To obtain the core flux, $S_{core}$, we seek through a large number of references and databases indicating to the core emissions, crosscheck these sources with the catalog given by Fan et al.(2011) and Pei et al.(2019), and choose those that were not listed in Fan et al.(2011) or Pei et al.(2019). And then we calculated their core-dominance parameters using Equation 2.

The data and their corresponding references are shown in Table 1. The complete Table for the whole sample is attached as an online material of this paper. Col. 1 gives the source names; Col. 2 classification (BL Lac: BL Lacertae object; FSRQ: flat spectrum radio quasar; Seyfert: Seyfert galaxy; G: galaxy; FRI&II: Fanaroff-Riley type I or II radio galaxy; U: unidentified sources); Col. 3 redshift, $z$; Col. 4 frequency in GHz for emission; Col. 5 core-emission in mJy; Col. 6 extended emission in mJy and Col. 7 total emission in mJy; Col. 8 references for Col 5, Col. 6 and Col. 7; Col. 9 core-dominance parameter at 5 GHz, log $R$; Col. 10 frequency in Ghz; Col. 11 total emission in mJy, data in Col. 10 and Col. 11 are from NED; Col. 12 the radio spectral index, $\alpha_R$ ($S_\nu \propto \nu^{-\alpha_R}$). Data in this table taken from Bal08: Balmaverde et al.(2008); Böc16: Böck et al.(2016); Bri94: Bridle et al.(1994); Bro15: Brotherton et al.(2015); CB07: Capetti & Balmaverde(2007); Cegł13: Cegłowski et al.(2013); Che15: Chen et al.(2015); Dro12: Drouart et al.(2012); Gio88: Giovannini et al.(1988); Han09: Hancock et al.(2009); JLG95: Johnson et al.(1995); Kel89: Kellermann et al.(1989); Kra17: Kravchenko et al.(2017); Lai83: Laing et al.(1983); LP95: Leahy & Perley(1995); LPR92: Liu et al.(1992); LPG02: Landt et al.(2002); Liu09: Liuzzo et al.(2009); Liu10: Liuzzo et al.(2010); Liu18: Liu et al.(2018); Man15: Mantovani et al.(2015); MA16: Marin & Antonucci(2016); Mül18: Müller et al.(2018); NRH95: Neff et al.(1995); OCU17: Odo et al.(2017); PT91: Perley & Taylor(1991); Raw90: Rawlings et al.(1990); RL15: Richards & Lister(2015); Ros06: Rossetti et al.(2006); Sar97: Saripalli et al.(1997); Smi16: Smith et al.(2016) Yu15: Yu et al.(2015) and Yua18: Yuan et al.(2018).

## 2.2 Estimated parameters

**Table 1** Sample for the whole sources

| Name | Class | $z$ | $\nu_1$ | $S_{core}$ | $S_{ext.}$ | $S_1^{Total}$ | Ref. | $\log R$ | $\nu_2$ | $S_2^{Total}$ | $\alpha$ |
|------|-------|-----|---------|------------|------------|---------------|------|----------|---------|---------------|----------|
| (1) | (2) | (3) | (4) | (5) | (6) | (7) | (8) | (9) | (10) | (11) | (12) |
| 1051+3911 | BL Lac | 1.372 | 5.0 | 72 | 0.42 | 72.42 | Che15 | 1.95 | 1.4 | 72.34 | 0.01 |
| 1429+400 | FSRQ | 0.276 | 5.0 | 118 | 8.14 | 126.14 | Che15 | 1.08 | 1.4 | 219 | 0.43 |
| 0122+015 | Seyfert | 1.559 | 5.0 | 47.2 | 88.66 | 135.86 | Man15 | -0.58 | 1.4 | 194 | 0.28 |
| 1117+592 | G | 0.159 | 5.0 | 4.80 | 48.96 | 53.76 | Man15 | -1.06 | 1.4 | 91.2 | 0.42 |
| 0930+369 | FRII | 2.395 | 5.0 | 0.29 | 79.61 | 70.90 | Dro12 | -2.78 | 1.4 | 284.1 | 1.09 |
| 0829+275 | U | 0.510 | 5.0 | 116 | 44.65 | 160.65 | Bro15 | 0.28 | 1.4 | 126.5 | -0.19 |
| . . . | . . . | . . . | . . . | . . . | . . . | . . . | . . . | . . . | . . . | . . . | . . . |

Notes: In this Table, Col. 1 gives the source names; Col. 2 classification (BL Lac: BL Lacertae object; FSRQ: flat spectrum radio quasar; Seyfert: Seyfert galaxy; G: galaxy; FRI&II: Fanaroff-Riley type I or II radio galaxy; U: unidentified sources); Col. 3 redshift, $z$; Col. 4 frequency in GHz for emission; Col. 5 core-emission in mJy; Col. 6 extended emission in mJy and Col. 7 total emission in mJy; Col. 8 references for Col 5, Col. 6 and Col. 7; Col. 9 core-dominance parameter at 5 GHz, $\log R$; Col. 10 frequency in Ghz; Col. 11 total emission in mJy, data in Col. 10 and Col. 11 are from NED; Col. 12 the radio spectral index, $\alpha_R$ ($S_\nu \propto \nu^{-\alpha_R}$). The complete Table for the whole sample is attached as an online material of this paper.

We found that $\log R$ are in the range from $-3.74$ to $3.43$ with an average value of $(\log R)|_{Total} = -0.06 \pm 1.09$ and a median of 0.25 for all the 966 sources. If we take the subclasses into account, we obtain that, correspondingly, from $-0.83$ to $3.10$ with an average value of $(\log R)|_{BL\ Lac} = 0.80 \pm 0.90$ and a median of 0.72 for 83 BL Lacs; from $-3.24$ to $3.43$ with an average value of $(\log R)|_{FSRQ} = 0.15 \pm 0.79$ and a median of 0.26 for 473 FSRQs; from $-2.25$ to $1.40$ with an average value of $(\log R)|_{Seyfert} = -0.09 \pm 0.71$ and a median of 0.003 for 101 Seyfert galaxies; from $-3.18$ to $2.77$ with an average value of $(\log R)|_{Galaxy} = -0.35 \pm 1.29$ and a median of $-0.52$ for 245 galaxies; from $-3.74$ to $0.01$ with an average value of $(\log R)|_{FRI\&II} = -1.85 \pm 0.91$ and a median of $-1.69$ for 52 FR type I and II radio galaxies; from $-2.87$ to $1.04$ with an average value of $(\log R)|_U = -0.33 \pm 1.34$ for 12 unidentified sources. Thus, for 563 blazars in our sample, an average value of $(\log R)|_{Blazar} = 0.25 \pm 0.84$ and a median of 0.27 from $-3.24$ to $3.43$ is also found accordingly (see Table 3).

Figure 1 shows the distribution of core-dominance parameter, $\log R$ (a) and the cumulative probability (b) for all the subclasses of our sample.

A Kolmogorov-Smirnov test (hereafter K-S test) test rejects the hypothesis that BL Lacs and quasars have the same parent distribution at $p = 1.01 \times 10^{-3}$ ($d_{max} = 0.46$). Likewise, we have $p = 5.63 \times 10^{-8}$ ($d_{max} = 0.32$) for FSRQs and Seyfert galaxies; $p = 2.90 \times 10^{-6}$ ($d_{max} = 0.30$) for Seyfert galaxies and galaxies; $p = 7.04 \times 10^{-11}$ ($d_{max} = 0.52$) for galaxies and FRI&II galaxies. Regarding to blazars and non-blazars, we obtain $p = 5.41 \times 10^{-35}$ ($d_{max} = 0.41$) (see Table 2).

Therefore, we found that the average core-dominance parameters for the sources follow the relation: $(\log R)|_{BL\ Lac} > (\log R)|_{FSRQ} > (\log R)|_{Seyfert} > (\log R)|_{Galaxy} > (\log R)|_{FRI\&II}$. Using the core-dominance parameter, blazars appear to be the most core-dominated population of AGNs.

We can evaluate 926 sources for their radio spectral indices ($\alpha_R$) in our whole sample, with an average value of $(\alpha_R)|_{Total} = 0.41 \pm 0.55$ and a median of 0.34, in the range from $-1.51$ to $2.53$. Besides, from $-1.48$ to $1.60$ with an average value of $(\alpha_R)|_{BL\ Lac} = 0.19 \pm 0.44$ and a median of 0.12 for 80 BL Lacs;

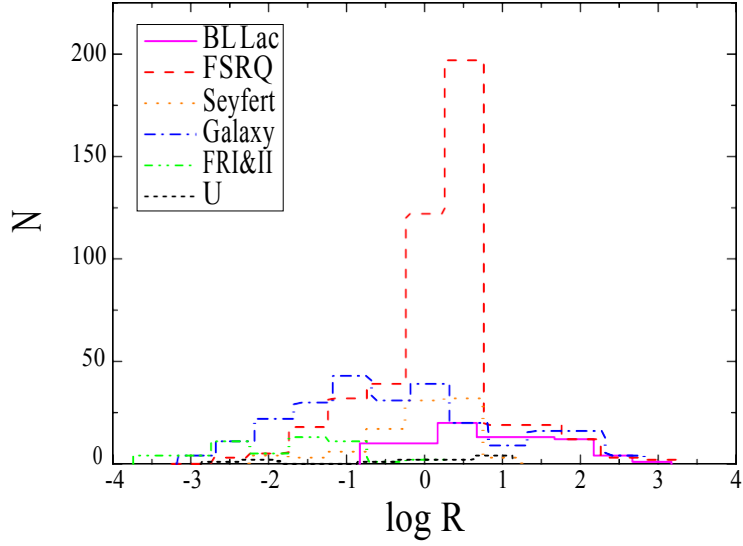

(a)

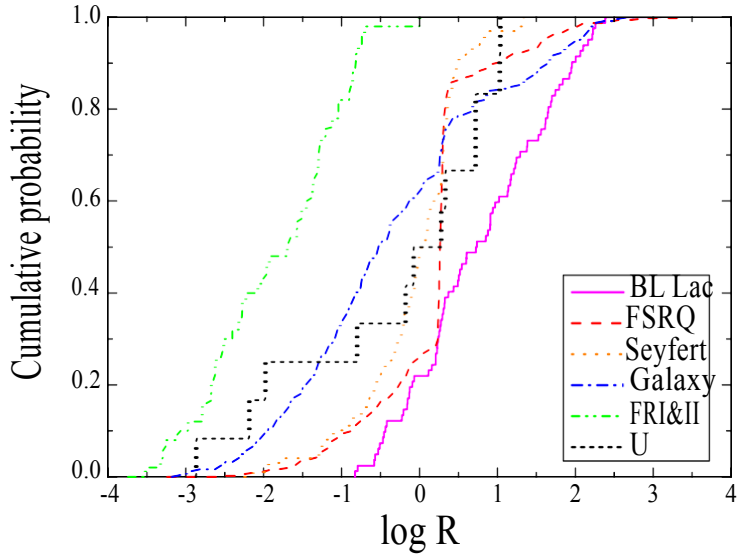

(b)

**Fig. 1** Distribution of core-dominance parameter, log $R$ (a) and the cumulative probability (b) for the whole sample. In this plot, *magenta solid line* refers for BL Lacs, *red dashed line* for FSRQs, *orange dotted line* for Seyferts, *blue dash-dotted line* for galaxies, *green dash-dott-dotted line* for FRIs&IIs and *black short dashed* line for unidentified sources, respectively.

from −1.25 to 2.53 with an average value of $(\alpha_R)|_{FSRQ} = 0.21 \pm 0.43$ and a median of 0.14 for 454 FSRQs; from −0.72 to 2.21 with an average value of $(\alpha_R)|_{Seyfert} = 0.77 \pm 0.66$ and a median of 0.80 for 95 Seyfert galaxies; from −1.32 to 2.01 with an average value of $(\alpha_R)|_{Galaxy} = 0.60 \pm 0.50$ and a median of 0.67 for 234 galaxies; from −0.43 to 2.28 with an average value of $(\alpha_R)|_{FRI\&IIs} = 1.03 \pm 0.35$ and a median of 1.01 for 52 FR I & II galaxies; from −1.51 to 1.19 with an average value of $(\alpha_R)|_U = 0.19 \pm 0.75$ for unidentified sources and from $-1.51$ to $2.53$ with an average value of $(\alpha_R)|_{Blazar} = 0.20 \pm 0.44$ for 541 blazar (see Table 3).

**Table 2** Statistical results for core-dominance parameter ($\log R$) in the whole sample

|          | Sample: A−B       | $N_A$ | $N_B$ | $d_{max}$ | $p$ |
|----------|-------------------|-------|-------|-----------|-----|
| $\log R$ | BL Lac−FSRQ       | 83    | 473   | 0.46      | $1.01 \times 10^{-3}$ |
|          | FSRQ−Seyfert      | 83    | 101   | 0.32      | $5.63 \times 10^{-6}$ |
|          | Seyfert−Galaxy    | 101   | 245   | 0.30      | $2.90 \times 10^{-6}$ |
|          | Galaxy−FRI&II     | 245   | 52    | 0.52      | $7.04 \times 10^{-11}$ |
|          | Blazar−non-Blazar | 563   | 403   | 0.41      | $5.41 \times 10^{-35}$ |
| $\alpha_R$ | BL Lac−FSRQ     | 80    | 454   | 0.20      | 1% |
|          | FSRQ−Galaxy       | 454   | 234   | 0.39      | $6.57 \times 10^{-21}$ |
|          | Galaxy−Seyfert    | 234   | 95    | 0.23      | $8.95 \times 10^{-4}$ |
|          | Seyfert−FRI&II    | 95    | 52    | 0.44      | $1.63 \times 10^{-6}$ |
|          | Blazar− non-Blazar| 541   | 385   | 0.47      | $1.27 \times 10^{-44}$ |

Figure 2 displays the distribution of radio spectral index, $\alpha_R$ (a) and the cumulative probability (b) for all the subclasses.

By adopting the K-S test, we obtain the following results: $p = 1\% (d_{max} = 0.20)$ for BL Lacs and FSRQs; $p = 6.57 \times 10^{-21}$ ($d_{max} = 0.39$) for FSRQs and galaxies; $p = 8.95 \times 10^{-4}$ ($d_{max} = 0.23$) for galaxies and Seyfert galaxies; $p = 1.63 \times 10^{-6}$ ($d_{max} = 0.44$) for Seyfert galaxies and FRI&II type galaxies and $p = 1.27 \times 10^{-44}$ ($d_{max} = 0.47$) for blazars and non-blazars (see Table 2).

We can find that the average radio spectral indices for all the sources follows a trend: $(\alpha_R)|_{FRI\&II} > (\alpha_R)|_{Seyfert} > (\alpha_R)|_{Galaxy} > (\alpha_R)|_{FSRQ} > (\alpha_R)|_{BL\ Lac}$, which is basically opposite to the trend in $\log R$. Through the K-S tests, we found that the distributions of $\log R$ and $\alpha_R$ in the various subclasses are almost significantly different, indicating that there are many different intrinsic properties among all the subclasses. However, considering the BL Lacs versus FSRQs with regard to $\alpha_R$, the result from K-S test demonstrates that there is no significant difference (with chance probability of $1.01 \times 10^{-3}$ for $\log R$ and of 1% for $\alpha_R$), which implies that, as the two subclasses of blazars, they hold some similar immanent properties.

For the radio core luminosity $\log L_{core}$ (W·Hz$^{-1}$), from $19.04$ to $28.33$ with an average value of $(\log L_{core})|_{Total} = 24.69 \pm 1.83$ and a median of 25.04 for the whole sample; from $20.81$ to $27.81$ with an average value of $(\log L_{core})|_{BL\ Lac} = 25.34 \pm 1.40$ and a median of 25.39 for BL Lacs; from $20.22$ to $28.33$ with an average value of $(\log L_{core})|_{FSRQ} = 25.64 \pm 1.17$ and a median of 25.58 for FSRQs; from $19.32$ to $26.64$ with an average value of $(\log L_{core})|_{Seyfert} = 22.30 \pm 1.87$ and a median of 21.61 for Seyferts; from $19.04$ to $26.98$ with an average value of $(\log L_{core})|_{galaxy} = 23.67 \pm 1.61$ and a median of 23.67 for galaxies; from $20.75$ to $27.05$ with an average value of $(\log L_{core})|_{FRI\&IIs} = 24.14 \pm 1.40$ and a median of 24.27 for FRI&IIs (also see Table 3).

On the other hand, for the radio extended luminosity $\log L_{ext.}$ (W·Hz$^{-1}$), we obtained that from $19.21$ to $28.57$ with an average value of $(\log L_{ext.})|_{Total} = 24.62 \pm 1.88$ and a median of 24.80 for all the sources; from $19.41$ to $28.02$ with an average value of $(\log L_{ext.})|_{BL\ Lac} = 24.44 \pm 1.82$ and a median of 24.52 for BL Lacs; from $19.50$ to $28.57$ with an average value of $(\log L_{ext.})|_{FSRQ} = 25.31 \pm 1.35$ and a median of

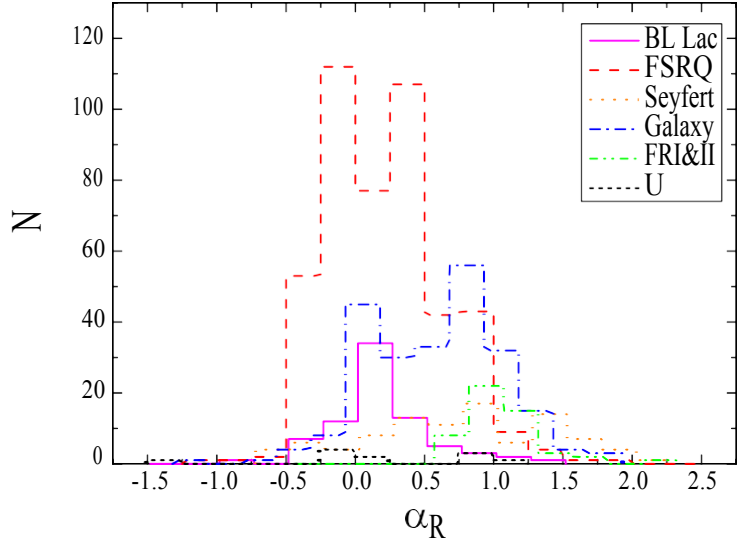

(a)

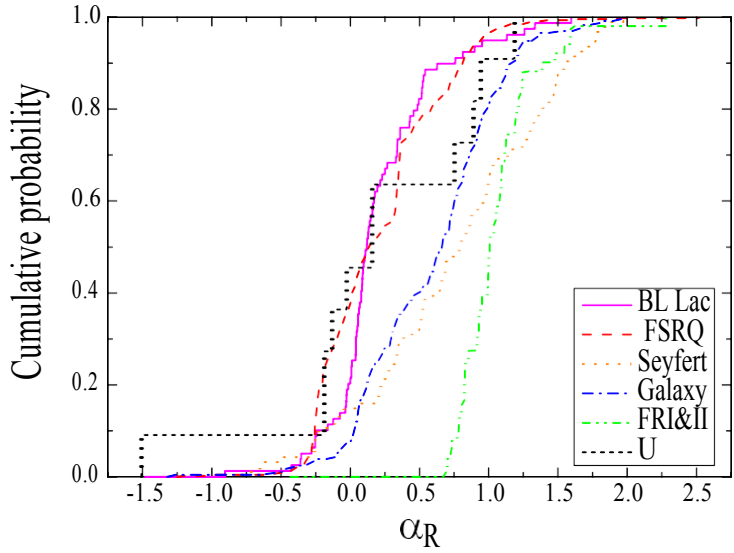

(b)

**Fig. 2** Distribution of radio spectral index, $\alpha_R$ (a) and the cumulative probability (b) for the whole sample. In this plot, the representations for all sources are the same as Figure 1.

25.19 for FSRQs; from $19.21$ to $27.22$ with an average value of $(\log L_{ext.})|_{Seyfert} = 22.35 \pm 1.88$ and a median of 21.63 for Seyferts; from $19.65$ to $27.87$ with an average value of $(\log L_{ext.})|_{galaxy} = 23.96 \pm 1.85$ and a median of 24.18 for galaxies; from $21.54$ to $27.84$ with an average value of $(\log L_{ext.})|_{FRI\&IIs} = 25.97 \pm 1.38$ and a median of 26. 36 for FRI&IIs (also see Table 3).

**Table 3** Average values for the whole sample

| Sample | $N$ | $(\log R)$ | $(\alpha_R)$ | $(\log L_{core})$ (W·Hz$^{-1}$) | $(\log L_{ext.})$ (W·Hz$^{-1}$) |
|---|---|---|---|---|---|
| Total | 966 | $-0.06 \pm 1.09$ | $0.41 \pm 0.55$ | $24.69 \pm 1.83$ | $24.62 \pm 1.88$ |
| BL Lac | 83 | $0.80 \pm 0.90$ | $0.19 \pm 0.44$ | $25.34 \pm 1.40$ | $24.44 \pm 1.82$ |
| FSRQ | 473 | $0.15 \pm 0.79$ | $0.21 \pm 0.43$ | $25.64 \pm 1.17$ | $25.31 \pm 1.35$ |
| Seyfert | 101 | $-0.09 \pm 1.40$ | $0.77 \pm 0.66$ | $22.30 \pm 1.87$ | $22.35 \pm 1.88$ |
| Galaxy | 245 | $-0.35 \pm 1.29$ | $0.60 \pm 0.50$ | $23.67 \pm 1.61$ | $23.96 \pm 1.85$ |
| FRI&II | 52 | $-1.85 \pm 0.91$ | $1.03 \pm 0.35$ | $24.14 \pm 1.40$ | $25.97 \pm 1.38$ |
| Unidentified | 12 | $-0.33 \pm 1.34$ | $0.19 \pm 0.75$ | $\cdots$ | $\cdots$ |
| Blazar | 563 | $0.25 \pm 0.84$ | $0.20 \pm 0.44$ | $\cdots$ | $\cdots$ |

## 3 DISCUSSION

From the previous studies, the core-dominance parameter, $R$, can act the role of the proxy of Doppler beaming effect and the orientation indicator of the jet(Urry & Padovani1995,Fan2003),

$$R(\theta) = f\gamma^{-n}[(1 - \beta\cos\theta)^{-n+\alpha} + (1 + \beta\cos\theta)^{-n+\alpha}], \tag{3}$$

where $f$ is the intrinsic ratio, $f = \frac{S_{core}^{in}}{S_{ext.}^{in}}$ (Fan & Zhang2003), $\theta$ the viewing angle, $\gamma$ the Lorentz factor, $\gamma = (1 - \beta^2)^{-1/2}$, $\beta \equiv c/v$, $\alpha$ is the radio spectral index and $n$ depends on the shape of the emitted spectrum and the physical detail of the jet, $n = 2$ for continuous jet and $n = 3$ for blobs. Therefore, $R$ is a good statistical measure and indicator of the relativistic beaming effect.

### 3.1 Correlation Analysis

We now turn our attention to linear correlation, if any, of the extended and core luminosity. In the two-component beaming model, the core emission are supposed to be the beamed component and the extended emission to be the unbeamed one. Because of the core-dominance parameter can take the indication of orientation, thus $\log R$ is also the indication of beaming effect. We use the extended luminosity at 5 GHz, $\log L_{ext.}$ in the unit of W·Hz$^{-1}$, to study the relationship between the beaming effect and unbeamed emission. Here, we K-corrected the flux densities, and then calculate the luminosity by $L_v = 4\pi d_L^2 S_v$, where $d_L$ is the luminosity distance in the unit of Mpc. We obatin the relation that $\log L_{ext.} = (0.84 \pm 0.02) \log L_{core} + (3.82 \pm 0.47)$ with a correlation coefficient $r = 0.82$ and a chance probability of $p \sim 0$ for the whole sample as shown in Figure3, which shows that the extended luminosity $\log L_{ext.}$ increases with increasing core luminosity $\log L_{core}$.

Adopting the two-component model, the core-dominance parameter, $R = L_{core}/L_{ext.}$, can be indicated as the form

$$1 + R = \frac{L_{core} + L_{ext.}}{L_{ext.}} = \frac{L_{total}}{L_{ext.}}, \tag{4}$$

If we take the assumption that the total luminosity of our compiled sample, $L_{total} = L_{core} + L_{ext.}$, is a constant, then we should expect that $R + 1$ is anti-correlated to $L_{ext.}$ when $R$ is much larger than unity, and there is no correlation between $R$ and $L_{ext.}$ when $R$ is much smaller than unity (Fan et al.2011;Pei et al.2019). A correlation between core-dominance parameter and extended luminosity is found for the whole sample (Figure 4), which displays that log $L_{ext.} = -(0.69 \pm 0.05) \log R + (24.58 \pm 0.06)$ with a correlation coefficient $r = -0.40$ and a chance probability of $p \sim 0$. In this plot, the representations of all symbols are the same as in Figure 3. This correlation implies that a lower extended luminosity source has a larger core-dominance parameter, $R$, showing either a large $f$ or a large $\delta$ since $R \propto f\delta^p$ (Ghiselini et al. 1993).

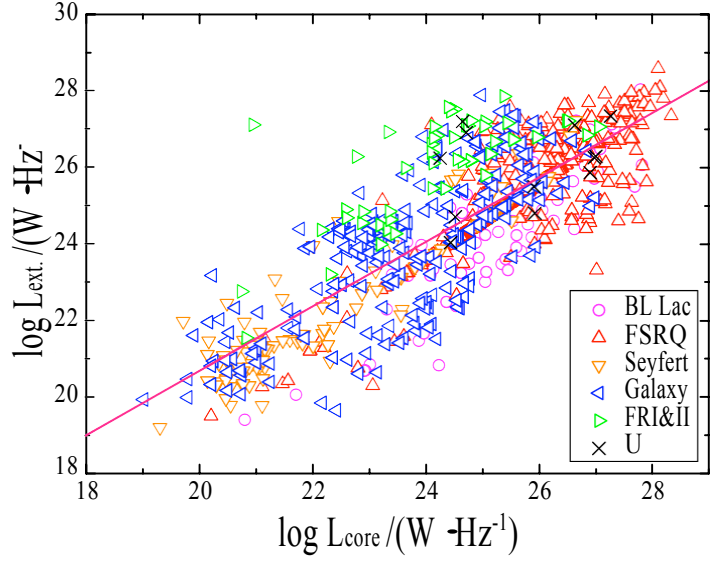

**Fig. 3** Plot of the extended luminosity (log $L_{ext.}$) against the core luminosity (log $L_{core}$) for the whole sample. Linear fitting gives that log $L_{ext.} = (0.84 \pm 0.02) \log L_{core} + (3.82 \pm 0.47)$ ($r = 0.82, p \sim 0$). In this plot, *magenta* § represents for BL Lac, *red* O for FSRQ, *orange* q for Seyfert, *blue* a for galaxy, *green* d for FRI & FRII and *black* × for unidentified sources.

The above correlation is perhaps from an evolutionary result. If the "young" AGNs have a powerful unremitting activity in the core or beamed component, however, the extended radiation has not yet had the enough time to accumulate, then those "young" sources have weak extended emission but strong beamed emission, resulting in a large log $R$ (Fan et al.2011).

## 3.2 Correlation between core-dominance parameter and radio spectral index

Based on the two component-model (Urry & Shafer1984),Fan et al.(2010) obtained a significative correlation between core-dominance parameter, $R$, and radio spectral index, $\alpha_R$

$$\alpha_{total} = \frac{R}{1+R}\alpha_{core} + \frac{1}{1+R}\alpha_{ext.}, \qquad (5)$$

where $\alpha_{total}$, $\alpha_{core}$ and $\alpha_{ext.}$ stands for total, core and extended components of radio spectral index $\alpha_R$, respectively. We take the spectral index ($\alpha_R$) that we calculated above is equal to the total component of spectral index for the sources, that is, $\alpha_R = \alpha_{total}$ under our consideration. We adopt this relationship to our sample and study the correlation between core-dominance parameter log $R$ and the total radio spectral index $\alpha_{\text{Total}}$, and obtain the theoretical derived fitting results of $\alpha_{\text{core}}$ and $\alpha_{\text{ext.}}$ by minimizing $\Sigma[\alpha_{\text{total}} - \alpha_{\text{core}} R/(1 + R) + \alpha_{\text{ext.}}(1 + R)]^2$ (also see Fan et al.2011; Pei et al.2019).

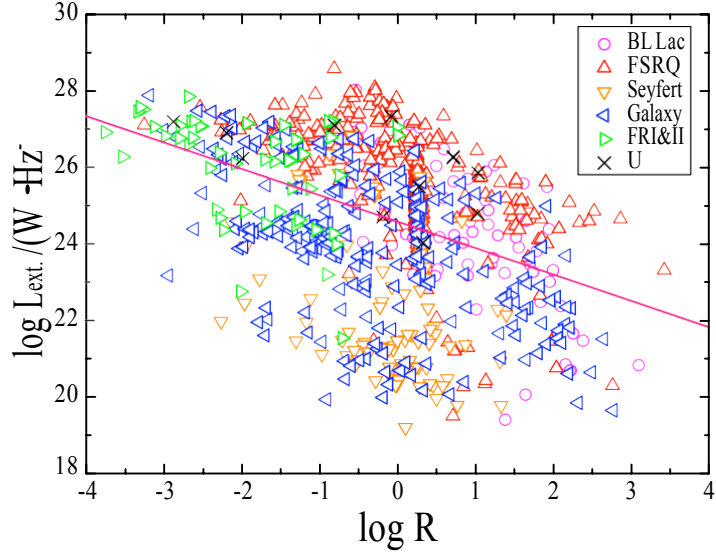

**Fig. 4** Plot of the extended luminosity ($\log L_{\text{ext.}}$) against the core-dominance parameter ($\log R$) for the whole sample. Linear fitting demonstrates that $\log L_{\text{ext.}} = -(0.69 \pm 0.05) \log R + (24.58 \pm 0.06)$ ($r = -0.40$, $p \sim 0$). In this plot, all representations of labels are the same as Figure 3.

The core-dominance parameters and radio spectral indices from Table 1 are plotted in Figure 5. When we adopt the above correlation in this current sample, $\alpha_{\text{core}} = -0.01 \pm 0.03$ and $\alpha_{\text{ext.}} = 0.83 \pm 0.03$ are obtained with a chance probability of $p \sim 0$ ($\chi^2 = 0.23$, $R^2 = 0.24$). The fitting result is shown in the Curve 1 in Figure 5. If we only consider the blazars, the derived fitting gives that $\alpha_{\text{core}} = -0.07 \pm 0.03$ and $\alpha_{\text{ext.}} = 0.59 \pm 0.04$ with a chance probability of $p \sim 0$ ($\chi^2 = 0.16$, $R^2 = 0.16$) (see Curve 2 in Figure 5). These derived results are consistent with the general consideration taking $\alpha_{\text{core}} = 0$ and $\alpha_{\text{ext.}} = 0.75$ (Fan et al.2011, also see Pei et al.2016,2019).

When we consider the subclasses separately, we can obtain a plot of the spectral index against the core-dominance parameter as shown in Figure 6(a) to (d) for BL Lacs, FSRQs, Seyfert galaxies and galaxies, respectively. The derived fitting results give that $\alpha_{\text{core}} = 0.03 \pm 0.06$ and $\alpha_{\text{ext.}} = 0.66 \pm 0.14$ with a chance probability of $p = 2.00 \times 10^{-4}$ ($\chi^2 = 0.01$, $R^2 = 0.13$) for BL Lacs, $\alpha_{\text{core}} = -0.09 \pm 0.04$ and $\alpha_{\text{ext.}} = 0.60 \pm 0.04$ with a chance probability of $p \sim 0$ ($\chi^2 = 0.16$, $R^2 = 0.17$) for FSRQs, $\alpha_{\text{core}} = 0.55 \pm 0.15$ and $\alpha_{\text{ext.}} = 0.98 \pm 0.14$ with a chance probability of $p = 1.23 \times 10^{-12}$ ($\chi^2 = 0.43$, $R^2 = 0.02$) for Seyferts, $\alpha_{\text{core}} = 0.22 \pm 0.06$ and $\alpha_{\text{ext.}} = 0.85 \pm 0.04$ with a chance probability of $p = \sim 0$ ($\chi^2 = 0.19$, $R^2 = 0.21$) for galaxies, respectively. However, for FRIs and FRIIs, we cannot get an appropriate fitting. All results are given in Table 4. The tendency for the spectral index to depend on the core-dominance parameter is probably due to the relativistic beaming effect and our fitting results imply that different subclasses show different degrees of relevance with regard to the beaming effect.

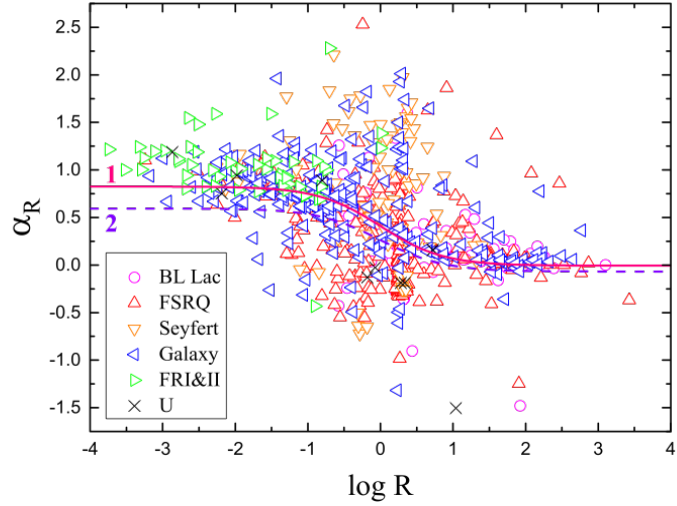

**Fig. 5** Plot of the the radio spectral index, $\alpha_R$, against the core-dominance parameter, $\log R$, for the whole sample. In this plot, all representations of labels are the same as Figure 3. Curve 1 in *solid pink* line corresponds to the derived fitting for all the sources as $\alpha_{core}$ (or $\alpha_j$)= $-0.01$ and $\alpha_{ext.}$ (or $\alpha_{unb}$)= $0.83$, curve 2 in *dash violet* line corresponds to the derived fitting for all the blazars as $\alpha_{core}$ = $-0.07$ and $\alpha_{ext.}$ = $0.59$.

**Table 4** The derived fitting results for radio spectral index against core-dominance parameter for the whole sample

| Sample | $\alpha_{core}$ | $\alpha_{ext.}$ | $\chi^2$ | $R^2$ | $p$ |
|--------|-----------------|-----------------|----------|-------|-----|
| Total | $-0.01 \pm 0.03$ | $0.83 \pm 0.03$ | 0.23 | 0.24 | $\sim 0$ |
| BL Lac | $0.03 \pm 0.06$ | $0.66 \pm 0.14$ | 0.17 | 0.13 | $2.00 \times 10^{-4}$ |
| FSRQ | $-0.09 \pm 0.04$ | $0.60 \pm 0.04$ | 0.16 | 0.17 | $\sim 0$ |
| Blazar | $-0.07 \pm 0.03$ | $0.595 \pm 0.04$ | 0.16 | 0.16 | $\sim 0$ |
| Seyfert | $0.55 \pm 0.15$ | $0.98 \pm 0.14$ | 0.43 | 0.02 | $1.23 \times 10^{-12}$ |
| Galaxy | $0.22 \pm 0.06$ | $0.85 \pm 0.04$ | 0.19 | 0.21 | $\sim 0$ |
| FRI&II | $\cdots$ | $\cdots$ | $\cdots$ | $\cdots$ | $\cdots$ |

In this paper, following Fan et al.(2011) and Pei et al.(2019), we obtain that $\log R|_{BL\ Lac} > \log R|_{FSRQ} > \log R|_{Seyfert} > \log R|_{Galaxy} > \log R|_{FRI\&FRII}$ averagely, and also roughly consistent with the distri- butions of their beaming factor values (e.g., Jorstad et al.2005; Richards & Lister2015; Sun et al.2015; Xue et al.2017). And for the circumstances of $\alpha_R$, we have the similar conclusion, on average, $\alpha_R|_{Blazar} < \alpha_R|_{Seyfert\swarrow Galaxy} < \alpha_R|_{FRI\&FRII}$, which is in accordance with our precious study (see Fan et al.2011; Pei

For the blazars in our current sample, we also found that the relation $(\log R)|_{\text{BL Lac}} > (\log R)|_{\text{FSRQ}}$ holds, which are semblable with the results drew in Fan et al.(2011) and Pei et al.(2019). Does that mean the beaming effect on BL Lacs is stronger than FSRQs? We do not think so. Fan(2003) studied the intrinsic

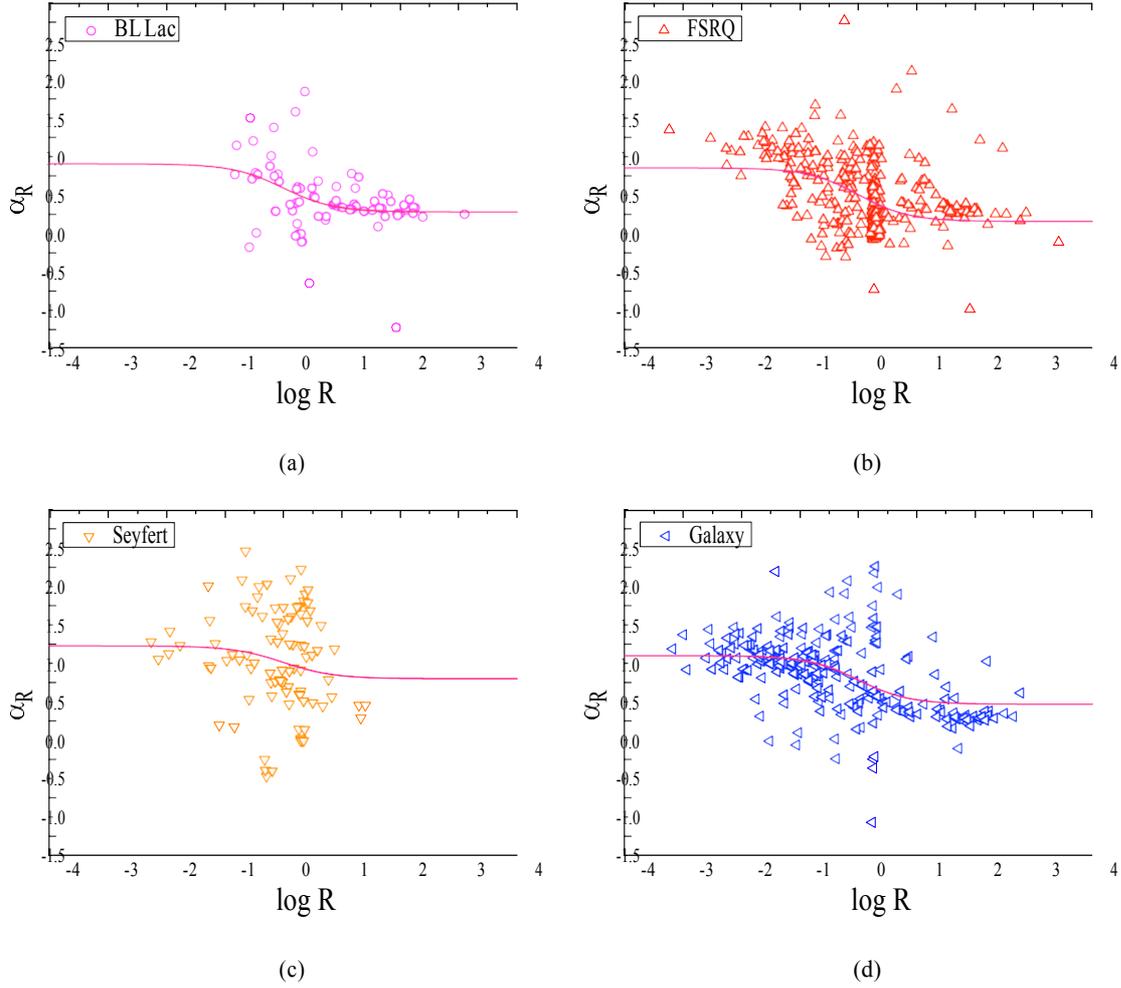

(a)

(b)

(c)

(d)

**Fig. 6** Plot of the radio spectral index, $\alpha_R$, against the core-dominance parameter, $\log R$, for BL Lacs (a), FSRQs (b), Seyfert galaxies (c) and galaxies (d). The derived fitting corresponds to $\alpha_{core}$ (or $\alpha_j$)= 0.03 and $\alpha_{ext.}$ (or $\alpha_{unb}$)= 0.66 for BL Lacs (a), $\alpha_{core}$ (or $\alpha_j$)= −0.09 and $\alpha_{ext.}$ (or $\alpha_{unb}$)= 0.60 for FSRQs (b), $\alpha_{core}$ (or $\alpha_j$)= 0.55 and $\alpha_{ext.}$ (or $\alpha_{unb}$)= 0.98 for Seyferts (c) and $\alpha_{core}$ (or $\alpha_j$)= 0.22 and $\alpha_{ext.}$ (or $\alpha_{unb}$)= 0.85 for galaxies (d), respectively.

ratio $f$ defined in Equation 3, and found that, $f$ of BL Lacs is on average higher than FSRQs, suggesting that the emissions from the jet are more dominant of BL Lacs from the unbeamed line emissions compared with FSRQs. And this result can also explain the differences in line emissions and polarizations between BL Lacs and FSRQs (Fan2003).

Previously many researchers have studied the unification model for BL Lacs ~ FRIs and quasars ~ FRIIs (e.g.Urry et al.1991;Xie et al.1993;Ubachukwu & Chukwude2002;Odo & Ubachukwu2013; Xue et al.2017), they all proposed that BL Lacs are unified with FRIs while quasars should be unified with FRIIs. Similar results are also obtained based on infrared (Fan et al.1997) and X-ray studies (Wang et al.2006). In the popular unification scenario by relativistic beaming effect, BL Lacs are believed to be the beamed counterparts of FRI radio galaxies, while quasars are believed to be the beamed counterparts of FRII radio galaxies (Urry & Padovani1995). However, we only have 32 FR type I radio galaxies and 20

**Table 5** Comparison of statistical results among Fan et al.(2011),Pei et al.(2019) and this work

| Sample | $N$ | | | $(\log R)$ | | | $(\alpha_R)$ | | | $\alpha_{core}$ | | | $\alpha_{ext.}$ | | |
|---|---|---|---|---|---|---|---|---|---|---|---|---|---|---|---|
| | Fan11 | Pei19 | TW | Fan11 | Pei19 | TW | Fan11 | Pei19 | TW | Fan11 | Pei19 | TW | Fan11 | Pei19 | TW |
| Total | 1223 | 2400 | 966 | $-0.35$ | $-0.34$ | $-0.06$ | 0.51 | 0.41 | 0.41 | $-0.07$ | $-0.08$ | $-0.01$ | 0.92 | 1.04 | 0.83 |
| BL Lac | 77 | 250 | 83 | | 0.87 | 0.55 | 0.80 | 0.16 | 0.22 | 0.19 | $-0.01$ | $-0.02$ | $-0.09$ | 0.65 | 0.70 | 0.66 |
| FSRQ | 495 | 520 | 473 | | 0.13 | 0.24 | 0.15 | 0.36 | 0.15 | 0.21 | $-0.09$ | $-0.34$ | $-0.09$ | 0.89 | 0.60 | 0.60 |
| Seyfert | 180 | 175 | 101 | $-0.39$ | $-0.37$ | $-0.09$ | 0.53 | 0.43 | 0.77 | $-0.01$ | $\cdots$ | 0.55 | 0.91 | $\cdots$ | 0.98 |
| Galaxy | 280 | 1178 | 245 | $-0.93$ | $-0.67$ | $-0.35$ | 0.73 | 0.57 | 0.60 | $-0.01$ | $-0.05$ | 0.22 | 0.91 | 0.88 | 0.85 |
| FRI&II | 119 | 153 | 52 | $-1.99$ | $-1.26$ | $-1.85$ | 0.94 | 0.63 | 1.03 | 0.34 | $-0.12$ | $\cdots$ | 0.97 | 1.04 | $\cdots$ |

Notes: Here, 'Fan11', 'Pei19' and 'TW' refers to Fan et al.(2011),Pei et al.(2019) and this work, respectively

type II galaxies in our latest sample, thus we do not proceed this unified scheme. The related studies can see our previous work (Fan et al.2011;Pei et al.2019).

The correlation between the core-dominance parameter $R$ and radio spectral index $\alpha_R$ and extragalactic radio sources is a significant study. Fan et al.(2011) and Pei et al.(2019) calculated the core-dominance parameters and the radio spectral indices for the compiled samples, and gave the relationship between $\alpha$ and $\log R$, indicating that $\alpha_R$ is associated with $\log R$. We also suggest that the relativistic beaming effect may result in an association between spectral index and core-dominance parameter for extragalactic sources in radio emission (also see Pei et al.2016). In the two-component beaming model, the relative prominence of the core regarding to the extended emission defined as the ratio of core-to-extended-flux density measured in the rest frame of the source $\log R$ has become a suitable statistical measure of orientation and good indicator of beaming effect.

In our previous work (Fan et al.2011;Pei et al.2019), we collected 1223 and 2400 AGNs, respectively. In this paper, we enlarge the AGNs sample and 966 sources are included. They are 83 BL Lacs, 473 FSRQs, 101 Seyferts, 245 galaxies, 52 FRIs & FRIIs and 12 unidentified sources (including 7 BCUs). We also discuss the relation between core-dominance parameter ($\log R$) and radio spectral index ($\alpha_R$) (see relation 5), and we obtain the similar relation and gain the derived fitting values for $\alpha_{core}$ and $\alpha_{ext.}$ as well. The comparison of previous work and this paper is shown in Table 5.

When all the sources in Fan et al.(2011),Pei et al.(2019) and this work are considered together, we own a large catalog with 4589 sources included relative to available $\log R$, and we have that $(\log R)|_{BL\ Lac} = 0.56 \pm 0.95$ in the range from $-3.18$ to 3.92 and a median of 0.35 for BL Lacs; from $-3.35$ to 3.44 with an average value of $(\log R)|_{FSRQ} = 0.14 \pm 0.94$ and a median of 0.24 for FSRQs; from $-3.20$ to 2.23 with an average value of $(\log R)|_{Seyfert} = -0.32 \pm 0.92$ and a median of $-0.29$ for Seyferts; from $-3.56$ to 3.08 with an average value of $(\log R)|_{Galaxy} = -0.72 \pm 0.99$ and a median of $-0.73$ for galaxies; from $-4.16$ to 1.27 with an average value of $(\log R)|_{FRI\&II} = -1.63 \pm 0.94$ and a median of $-1.46$ for FRIs&FRIIs. The K-S test demonstrates that $p = 3.08 \times 10^{-12}$ ($d_{max} = 0.20$) for BL Lacs and FSRQs; $p = 1.09 \times 10^{-26}$ ($d_{max} = 0.28$) for FSRQs and Seyferts; $p = 6.46 \times 10^{-25}$ ($d_{max} = 0.27$) for Seyferts and galaxies; $p = 5.30 \times 10^{-41}$ ($d_{max} = 0.40$) for galaxies and FRIs&FRIIs. Therefore, we can draw our conclusion that there is a sequence for core-dominance parameter in the subclasses of AGNs: $\log R|_{BL\ Lac} > \log R|_{FSRQ} > \log R|_{Seyfert} > \log R|_{Galaxy} > \log R|_{FRI\&II}$, averagely, which it is associated with the beaming model.

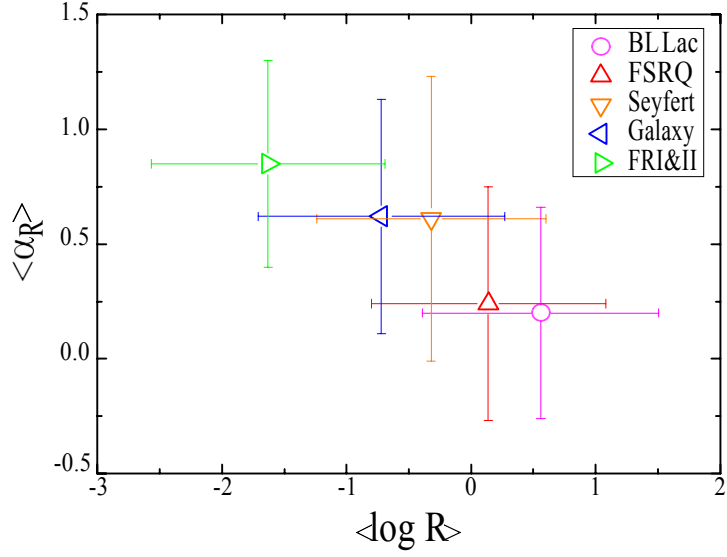

**Fig. 7** Plot of the average values of radio spectral indices ($(\alpha_R)$) against core-dominance parameters ($(\log R)$) with error bar for this work together with the previous two samples (Fan et al. 2011;Pei et al.2019). In this plot, all representations of labels are the same as Figure 3.

For $\alpha_R$, we also obtain that $(\alpha_R)|_{BL\,Lac} = 0.20 \pm 0.46$ in the range from $-1.88$ to $1.60$ and a median of $0.17$ for BL Lacs; from $-1.66$ to $2.53$ with an average value of $(\alpha_R)|_{FSRQ} = 0.24 \pm 0.51$ and a median of $0.23$ for FSRQs; from $-2.42$ to $2.21$ with an average value of $(\alpha_R)|_{Seyfert} = 0.61 \pm 0.62$ and a median of $0.68$ for Seyferts; from $-2.31$ to $2.12$ with an average value of $(\alpha_R)|_{Galaxy} = 0.62 \pm 0.51$ and a median of $0.74$ for galaxies; from $-1.98$ to $2.28$ with an average value of $(\alpha_R)|_{FRI\&FRII} = 0.85 \pm 0.45$ and a median of $0.92$ for FRIs&FRIIs. And the K-S test shows that $p = 3.79 \times 10^{-4}$ ($d_{max} = 0.11$) for BL Lacs and FSRQs; $p = 5.79 \times 10^{-94}$ ($d_{max} = 0.38$) for FSRQs and Seyferts/galaxies; $p = 6.60 \times 10^{-25}$ ($d_{max} = 0.33$) for Seyferts/galaxies and FRIs&FRIIs. We also reveal a sequence for radio spectral index in the subclasses: on average, $\alpha_R|_{BL\,Lac} < \alpha_R|_{FSRQ} < \alpha_R|_{Seyfert/Galaxy} < \alpha_R|_{FRI\&II}$. These two sequences elucidate the relationship between $\log R$ and $\alpha_R$ extends over all the sources in the large sample combined with Fan et al. (2011),Pei et al.(2019) and this work. Figure 7 displays the plot of the averaged of radio spectral indices ($(\alpha_R)$) against core-dominance parameters ($(\log R)$) for the whole sample together with the previous two samples.

According to the relation 5, in the relativistic beaming scenario for highly beamed sources, we have $\log R \gg 0$, which leads to $\alpha_{total} \approx \alpha_{core}$. And $\alpha_R$ is dominated by $\alpha_{ext.}$ for the case $\log R \ll 0$. Therefore, we can consider that the association between core-dominance parameters and spectral indices may suggest that relativistic beaming could influence the spectral characteristics of this extreme class of objects.

# 4  CONCLUSION

From our discussions, given the the core-dominance parameters, $\log R$ and radio spectral indices, $\alpha_R$, the $\alpha_{core}$ and $\alpha_{ext.}$ can be obtained. In this paper, we compiled 966 objects with the relevant data to calculate the core-dominance parameters. We enlarge the AGNs sample with the available core-dominance parameters and we also make further statistical analysis. A larger catalog has been compiled, combining with our previous papers (Fan et al.2011;Pei et al.2019). These samples with core-dominance parameters provide us a significant implement to shed new light on the fascinating aspects of AGNs, e.g. blazars. Now we draw the following conclusions:

1. Core-dominance parameters ($\log R$) are quite different for different subclasses of AGNs: on average, the following sequence holds: $\log R|_{\text{BL Lac}} > \log R|_{\text{FSRQ}} > \log R|_{\text{Seyfert}} > \log R|_{\text{Galaxy}} > \log R|_{\text{FRI\&II}}$.

2. A theoretical correlation fitting between core-dominance parameter ($\log R$) and radio spectral index ($\alpha_R$) is adopted and also obtained for all subclasses, which indicates the radio spectral index is dependent on the core-dominance parameter, probably from the relativistic beaming effect. And $\alpha_{core} = -0.01$ and $\alpha_{ext.} = 0.83$ are obtained in this work.

3. There is an anti-correlation between extended-luminosity ($\log L_{ext.}$) and core-dominance parameter in different kinds of objects.

**Acknowledgements** This work is partially supported by the National Natural Science Foundation of China (11733001, U1531245), Natural Science Foundation of Guangdong Province (2017A030313011), and supports for Astrophysics Key Subjects of Guangdong Province and Guangzhou City.

## References

Abdo, A. A., Ackermann, M., & Ajello, M. 2010, ApJ, 723, 1082 2

Aller, H. D., Aller, M. F., Plotkin, R. M., Hughes, P. A., & Hodge, P. E. 2003, in Bulletin of the American Astronomical Society, Vol. 35, American Astronomical Society Meeting Abstracts, 1310 2

Andruchow, I., Romero, G. E., & Cellone, S. A. 2005, A&A, 442, 97 2

Balmaverde, B., Baldi, R. D., & Capetti, A. 2008, A&A, 486, 119 4

Böck, M., Kadler, M., Müller, C., et al. 2016, A&A, 590, A40 4

Bridle, A. H., Hough, D. H., Lonsdale, C. J., Burns, J. O., & Laing, R. A. 1994, AJ, 108, 766 4

Brotherton, M. S., Singh, V., & Runnoe, J. 2015, MNRAS, 454, 3864 4

Capetti, A., & Balmaverde, B. 2007, A&A, 469, 75 4

Cegłowski, M., Gawroński, M. P., & Kunert-Bajraszewska, M. 2013, A&A, 557, A75 4

Cellone, S. A., Romero, G. E., & Araudo, A. T. 2007, MNRAS, 374, 357 2

Chen, Y. Y., Zhang, X., Zhang, H. J., & Yu, X. L. 2015, MNRAS, 451, 4193 4

Drouart, G., De Breuck, C., Vernet, J., et al. 2012, A&A, 548, A45 4

Fan, J. H. 2003, ApJ, 585, L23 2,8,12

Fan, J. H. 2005, A&A, 436, 799 2

Fan, J. H., Cheng, K. S., Zhang, L., & Liu, C. H. 1997, A&A, 327, 947 2,13

Fan, J.-H., Yang, J.-H., Pan, J., & Hua, T.-X. 2011, Research in Astronomy and Astrophysics, 11, 1413 2, 3,4,9,10,11,12,13,14,15

Fan, J.-H., Yang, J.-H., Tao, J., Huang, Y., & Liu, Y. 2010, PASJ, 62, 211 10

Fan, J. H., & Zhang, J. S. 2003, A&A, 407, 899 2, 9

Fanaroff, B. L., & Riley, J. M. 1974, MNRAS, 167, 31P 2

Ghisellini, G., Padovani, P., Celotti, A., & Maraschi, L. 1993, ApJ, 407, 65 10

Giovannini, G., Feretti, L., Gregorini, L., & Parma, P. 1988, A&A, 199, 73 4

Hancock, P. J., Tingay, S. J., Sadler, E. M., Phillips, C., & Deller, A. T. 2009, MNRAS, 397, 2030 4

Johnson, R. A., Leahy, J. P., & Garrington, S. T. 1995, MNRAS, 273, 877 4

Jorstad, S. G., Marscher, A. P., Lister, M. L., et al. 2005, AJ, 130, 1418 11

Kellermann, K. I., Sramek, R., Schmidt, M., Shaffer, D. B., & Green, R. 1989, AJ, 98, 1195 4

Kravchenko, E. V., Kovalev, Y. Y., & Sokolovsky, K. V. 2017, MNRAS, 467, 83 4

Laing, R. A., Riley, J. M., & Longair, M. S. 1983, MNRAS, 204, 151 4

Landt, H., Padovani, P., & Giommi, P. 2002, MNRAS, 336, 945 4

Leahy, J. P., & Perley, R. A. 1995, MNRAS, 277, 1097 4

Lin, C., & Fan, J.-H. 2018, Research in Astronomy and Astrophysics, 18, 120 2

Liu, J., Bignall, H., Krichbaum, T., et al. 2018, Galaxies, 6, 49 4

Liu, R., Pooley, G., & Riley, J. M. 1992, MNRAS, 257, 545 4

Liuzzo, E., Giovannini, G., Giroletti, M., & Taylor, G. B. 2009, A&A, 505, 509 4

Liuzzo, E., Giovannini, G., Giroletti, M., & Taylor, G. B. 2010, A&A, 516, A1 4

Mantovani, F., Bondi, M., Mack, K. H., et al. 2015, A&A, 577, A36 4

Marin, F., & Antonucci, R. 2016, ApJ, 830, 82 4

Müller, C., Kadler, M., Ojha, R., et al. 2018, A&A, 610, A1 4 Neff,

S. G., Roberts, L., & Hutchings, J. B. 1995, ApJS, 99, 349 4

Odo, F. C., Chukwude, A. E., & Ubachukwu, A. A. 2017, Ap&SS, 362, 23 4

Odo, F. C., & Ubachukwu, A. A. 2013, Ap&SS, 347, 357 13

Orr, M. J. L., & Browne, I. W. A. 1982, MNRAS, 200, 1067 2

Pei, Z.-Y., Fan, J.-H., Bastieri, D., Sawangwit, U., & Yang, J.-H. 2019, Research in Astronomy and
    Astrophysics, 19, 0702 3, 4, 9, 10, 11, 12, 13, 14, 15

Pei, Z. Y., Fan, J. H., Liu, Y., et al. 2016, Ap&SS, 361, 237 2, 3, 11, 14

Perley, R. A., & Taylor, G. B. 1991, AJ, 101, 1623 4

Rawlings, S., Saunders, R., Miller, P., Jones, M. E., & Eales, S. A. 1990, MNRAS, 246, 21P 4

Richards, J. L., & Lister, M. L. 2015, ApJ, 800, L8 4, 11

Romero, G. E., Cellone, S. A., & Combi, J. A. 2000, A&A, 360, L47 2

Romero, G. E., Cellone, S. A., Combi, J. A., & Andruchow, I. 2002, A&A, 390, 431 2

Rossetti, A., Fanti, C., Fanti, R., Dallacasa, D., & Stanghellini, C. 2006, A&A, 449, 49 4

Saripalli, L., Patnaik, A. R., Porcas, R. W., & Graham, D. A. 1997, A&A, 328, 78 4

Smith, K. L., Mushotzky, R. F., Vogel, S., Shimizu, T. T., & Miller, N. 2016, ApJ, 832, 163 4

Sun, X.-N., Zhang, J., Lin, D.-B., et al. 2015, ApJ, 798, 43 11

The Fermi-LAT collaboration. 2019a, arXiv e-prints, arXiv:1902.10045 2, 3

The Fermi-LAT collaboration. 2019b, arXiv e-prints, arXiv:1905.10771 2

Ubachukwu, A. A., & Chukwude, A. E. 2002, Journal of Astrophysics and Astronomy, 23, 235 13

Urry, C. M., & Padovani, P. 1995, PASP, 107, 803 2, 8, 13

Urry, C. M., Padovani, P., & Stickel, M. 1991, ApJ, 382, 501 13

Urry, C. M., & Shafer, R. A. 1984, ApJ, 280, 569 2, 10

Wang, Y.-X., Zhou, J.-L., Yuan, Y.-H., Chen, J.-L., & Yang, J.-H. 2006, Chinese Journal of Astronomy and Astrophysics Supplement, 6, 357 13

Wills, B. J., Wills, D., Breger, M., Antonucci, R. R. J., & Barvainis, R. 1992, ApJ, 398, 454 2

Xie, G. Z., Liu, H. T., Cha, G. W., et al. 2005, AJ, 130, 2506 2

Xie, G. Z., Zhang, Y. H., Fan, J. H., & Liu, F. K. 1993, A&A, 278, 6 13

Xue, Z.-W., Zhang, J., Cui, W., Liang, E.-W., & Zhang, S.-N. 2017, Research in Astronomy and Astrophysics, 17, 090 11, 13

Yang, J., Fan, J., Liu, Y., et al. 2018a, Science China Physics, Mechanics, and Astronomy, 61, 59511 2

Yang, J. H., Fan, J. H., Zhang, Y.I., et al. 2018b, Acta Astronomica Sinica, 59, 38 2

Yu, X., Zhang, X., Zhang, H., et al. 2015, Ap&SS, 357, 14 4

Yuan, Z., Wang, J., Worrall, D. M., Zhang, B.-B., & Mao, J. 2018, ApJS, 239, 33 4

Zhang, Y.-W., & Fan, J.-H. 2008, ChJAA (Chin. J. Astron. Astrophys.), 8, 385 2